KIEV NATIONAL TARAS SHEVCHENKO UNIVERSITY
PHYSICS DEPARTMENT

Andrey Alexandrovich SEMENOV

# PHASE SPACE LOCALIZATION OF A SCALAR CHARGED PARTICLE

01.04.02 – theoretical physics

**REVIEW**
of the thesis for the degree of
Doctor of Philosophy (Candidate of Physics and Mathematics)

Kiev – 2002

The thesis is a manuscript.

The work has been performed in Institute of Physics of National Academy of Sciences of Ukraine.

Scientific leader — Doctor of Sciences (Physics and Mathematics),<br>Senior Researcher<br>Lev Bohdan Ivanovich.<br>Institute of Physics, NAS of Ukraine,<br>Leading Researcher.

Official referees:

Doctor of Sciences (Physics and Mathematics), Senior Researcher Sytenko Yuriy Oleksiyovich, Head of Department of Nuclear Theory and Quantum Field Theory, N. N. Bogolyubov Institute for Theoretical Physics, NAS of Ukraine,

Doctor of Philosophy (Candidate of Physics and Mathematics), Senior Researcher Golod Petro Ivanovich, Head of Department of Physics and Mathematics, National University "Kievan Mohyla Academy".

Leading institution:

NSC "Kharkov Institute of Physics and Technology", A.I. Akhiezer Institute for Theoretical Physics.

The defense will be held ________________2002 at _____ at the session of Specialiezed Scientific Council D 26.001.08 of Kiev National Taras Shevchenko University, address: 6 Academician Glushkov pr., Kiev 03127, Ukraine.

The thesis is available at the library of Kiev National Taras Shevchenko University at address: 66 Volodymirska St., Kiev, Ukraine.

Review has been sent ________________2002.

Scientific Secretary of Specialized<br>Scientific Council, Doctor of Philosophy (Candidate of Physics and Mathematics) — O.S. Svechnikova

# GENERAL CHARACTERISTIC OF THE WORK.

*Actuality.* Covariant (four-dimensional) representations are nowadays very popular among the existing relativistic generalizations of the Wigner function formalism. Some of them have very serious disadvantages (e.g., divergence of the integrals in the Wigner function definitions); moreover, these approaches do not take into account the reference frame where the quantum state reduction takes place.

The matrix-valued distributions that constitute an important class of three-dimensional approaches do not have this problem. However, the question as to the influence of a non-trivial charge structure of the position operator on the motion in the phase space has not been studied sufficiently in the framework of this approach.

Relativistic coherent states have been considered by many authors from different points of view. Malkin and Man'ko [JETP, - 1969, Vol. 28, P. 527] have introduced the coherent states for spin ½ and spin 0 particles in a constant homogeneous magnetic field. These states satisfy the charge superselection rule, but the position and momentum means are not connected with the real and imaginary parts of the coherent state parameter. In fact, these states corresponds to the approximation of non-local theory.

In the works by Bagrov, Gitman *et al.* (see, e.g., [Bagrov V.G., Baldiotti M.C., Gitman D.M., Shirokov I. V. New solutions of relativistic wave equations in magnetic fields and longitudinal fields: Preprint / LANL: hep-th/0110037]), the motion of a scalar charged particle (including the case of a constant homogeneous magnetic field) has been considered using the zero-plane concept (light cone variables). It makes possible to represent the Klein-Gordon equation in the form of the Schrodinger equation for a harmonic oscillator and to get the relevant coherent states. The position and momentum means coincide with the classical solutions in this approach. However, from the author point, this encounters some interpretation difficulties.

The thesis is devoted to problems arising in the context of the above mentioned theories. The importance of such studies derives from its fundamental role in the quantum field theory, in understanding of the quantum measurement, and, finally, in having a simple analogy in the solid state physics and its potential applied aspect in future quantum technologies.

*Connection with scientific programs, plans, themes.* The work has been carried out in the framework of the following State Budjet projects:

1. "Collective processes in small and mesophase systems", code 1.4.1. V/42, State registration number 0198u001418.

2. "Kinetic, electric and optic properties of small systems", code 1.4.1. V/67, State registration number 0101u000353.

Working on these projects, the author has studied influence of the band structure of the energy spectrum on the motion of a particle in the phase space representation using, as an example, the simplest model, namely, the model of a scalar charged particle.

*Aim and tasks of the investigation.* The aim of this work is to find peculiarities of strongly localized states of a quantum particle that can arise because of a non-trivial charge structure of the position and momentum operators and effective non-locality of the Hamiltonian (existence of the derivatives up to the infinity order). To this end one needs to consider the following tasks:

1. To determine influence of a constant homogeneous magnetic field on the even (observable) parts of the position and momentum operators and on the relevant commutation relations.

2. To construct the matrix-valued Weyl—Wigner—Moyal (WWM) formalism for scalar charged particles.

3. To construct the WWM formalism for the special class of observables that are arbitrary combinations of the position and momentum (charge-invariant observables).

4. To construct relativistic coherent states that both take into account the non-trivial charge structure of the position and momentum operators and satisfy the charge superselection rule.

5. To determine influence of the physical vacuum on the means of dynamical variables under the condition that no particle pair is created, using, for an example, the first and second moments of the position and momentum.

6. To consider entangled coherent states and determine how strongly the localization influences their properties.

The object of the investigation: the influence of the filled negative values energy band of relativistic particles (Dirak Sea) and very strong localization of a quantum state on the observables.

The subject of investigation: strongly localized states of spineless particles in the phase space representation.

Methods of the investigation:

1. The general method used in the thesis is the WWM representation. This is generalized for the relativistic case taking into account the non-trivial charge structure of the position and momentum operators.

2. The transform to the phase space representation is very useful in the case when the starting equation is in the form of the Schrodinger equation. To this end the Feshbach—Villars formalism is used in the work.

3. Some physical processes are investigated using coherent states. This method should be adopted to the relativistic case.

*Scientific novelty of the obtained results:*

1. The relativistic coherent states that both take into account the non-trivial charge structure of the position and momentum operators and satisfy the charge superselection rule have been constructed.

2. It has been shown that even and odd parts of the operators of observables which are arbitrary combinations of the position and momentum (charge-invariant observables) are uniquely related to each other. The relevant expression has been found.

3. It has been shown that in the case of a particle in a constant homogeneous magnetic field the evolution equations for charge-invariant observables coincide with their analogues in the non-local theory. The differences between the theories are in the class of functions that represent non-stationary states of the system.

4. A low frequency modulation (additional damping) of the orbit radius of the particle rotation in a constant homogeneous magnetic field has been found.

5. It has been shown that the “mean position” operators of the rotational motion for a particle in a constant homogeneous magnetic field generate the deformed Heisenberg—Weyl algebra.

6. It has been shown that the energy of the entangled coherent states of two fermions is smaller than that in the non-relativistic case.

*Practical importance of the obtained results.* The results obtained in the thesis can be used in experimental investigations of the one-particle processes in relativistic systems, and in other systems with the band structure of the energy spectrum (e.g., solids). They can be considered as a basis for developing the relativistic quantum tomography method and tomography of the conduction electrons in solids.

The effective increase in the coherence between eigenstates of the Hamiltonian can be used in the experimental investigations of a decoherence process because this allows one to consider dynamics of the interference terms in a more detail. Such an effect can be useful in applications when the quantum character of information plays crucial role (e.g., in quantum computers).

Low frequency modulation (additional damping) of the orbit radius of a particle in strong magnetic fields can be a subject of astrophysical investigations. Observation of this effect (e.g., on the synchrotron radiation from neutron stars) would, first of all, indicate the reference frame where reduction of the quantum state takes place, and, second, give an additional possibility for the magnetic field estimation.

*Personal contribution of the author.* The work has been carried out in the collaboration with Leading Researcher, Doctor of Sciences (Physics and Mathematics) B.I. Lev (Institute of Physics, NAS of Ukraine) and Associated Professor, Doctor of Philosophy (Candidate of Physics and Mathematics) C.V. Usenko (Kiev National Taras Shevchenko University). The author has actively participated at all stages of the investigation, from formulating the problems to finding the solutions and presenting the results .

The general idea and most of the calculations in the works [1,3,4,5] belong to the author. In the work [2], the idea to use the secondary quantization in the basis of displacement number states, to calculate the means of the energy (without their analysis) in the non-relativistic case, and consideration of the relativistic case belong to the author.

The valuable advice, regarding estimation of the functional factorial and existence of the resolution of unity for the non-linear coherent states of relativistic rotator, were given by Professor J.R. Klauder (University of Florida, USA), Professor K.A. Penson, and Doctor J.-M. Sixdeniers (Universite Pierre et Marie Curie, Paris, France).

*Approbation of the work results.*

1. Workshops of the Theoretical Physics Department, Institute of Physics, NAS of Ukraine.

2. International Workshop "Mathematical Physics – today, Priority Technologies – for tomorrow", Kiev, Ukraine, 12—17 May 1997.

3. Fifth International Conference on Squeezed States and Uncertainty Relations. Balatonfured, Hungary, 27—31 May 1997.

4. Eighth Ukrainian Conference and School "Plasma Physics and Controlled Fusion". Alushta, Crimea, 11—16 September 2000.

5. Seventh International Conference on Squeezed States and Uncertainty Relations. Boston, Massachusetts, USA, 4—7 June 2001.

6. Seventh International Wigner Symposium. Baltimore, Maryland, USA, 24—29 August 2001.

*Publications.* The results of the thesis have been published in four articles [1,2,3,4] and in three conferences proceedings [5,6,7].

*Structure and volume of the thesis.* The thesis consists of Introduction, seven Chapters, and Conclusions with the review of general results. This has a volume of 174 pages including 1 Appendix (3 pages), 10 figures (10 pages), and 1 table (1 page). The Bibliography includes 142 References.

## GENERAL CONTENTS

*INTRODUCTION* of the work is devoted to description of the current state of the scientific problem and the reasons why the formalism of phase space representation has not been sufficiently developed in the relativistic case; namely, these are the existence of some conceptual problems and impossibility (until very recently) of experimental observation of a quantum information processes with relativistic values of parameters.

In INTRODUCTION, also the actuality of the theme of the thesis is justified, as well as the aim of the investigation, scientific novelty, and practical importance of the obtained results are described.

*FIRST CHAPTER*, "Review of the literature and field of research", is devoted to description of the general statements of non-relativistic representation on the phase-space (WWM formalism, coherent states and their applications), problems of generalization to the relativistic case, and different approaches to it.

Section 1 describes the non-relativistic WWM formalism. It is based on the Weyl transformation from operators to symbols (functions on the phase space). In this representation the main difference between quantum and classical mechanics can be reduced to redefinition of the operations of the usual product and Poisson bracket. The non-commutative operation of the star-product appears in quantum mechanics instead of the first one, and the Moyal bracket appears instead of the second one.

A very important difference between quantum and classical mechanics is in the specific condition on the class of functions that can represent real states of physical systems. This means that any function (even normalized) cannot be considered as a Wigner function.

Section 2 describes the formalism of coherent states and their generalizations. Except of standard coherent states, the so-called non-linear coherent states are considered in detail. They can be defined as eigenstates of the deformed annihilation operator $\hat{b}_f$ which is defined by means of the standard annihilation operator $\hat{b}$, and the deformed function $f$ in the following way:

$$\hat{b}_f = \hat{b} f\left(\hat{b}^{+}\hat{b}\right).$$

Such states naturally appear in the consideration of particles in laser traps. In other words, they describe quite well the motion of a particle in a field of a potential (non-quadratic).

Section 3 describes applications of the formalism which can be interesting in the context of the thesis. Namely, the quantum tomography method, entangled coherent states, and applications in quantum kinetics are discussed.

The conceptual problems which make difficult a correct generalization of the phase space representation formalism for relativistic case are considered in section 4. Two of these problems are emphasized:

1. The Weyl transformation (similar to another transformation from operator to symbol) is not Lorentz invariant. This means that it does not include (and cannot include) time as an independent dynamical variable.

2. A well defined position operator does not exist in relativistic quantum mechanics. This manifests itself in the fact that, on the one hand, an eigenfunction of the standard position operator is a superposition of states with different signs of charge, and, on another hand, the Newton—Wigner position operator is not well-defined in respect to the Lorentz invariance.

The existing experimental data and theoretical works are analyzed in detail. This makes possible to formulate the input statements of the thesis:

1. The standard (three-dimensional) Weyl transformation is used. At that, equations are written down in the reference frame, where the reduction of the quantum state takes place.

2. It is assumed that the position operator has a non-trivial charge structure. Results, whenever it is possible, are compared with the non-local theory to find the consequences that may be verified in experiment.

Hereinafter, a variant of relativistic quantum mechanics with the Hamiltonian of the form

$$\hat{H} = \sqrt{m^2c^4 + c^2(\hat{\pi} - eA(\xi))^2} + e\varphi(\xi),$$

is referred to as a non-local theory.

Different approaches to relativistic generalization of the phase space representation developed by other authors are considered in section 5.

*SECOND CHAPTER* “Charge structure of the position and momentum operators” is devoted to consideration of what is the form of the position and momentum operators in the representation of non-local theory (Feshbach—Villars in the case of a free particle). Taking into account the superselection rule and well-known results by Feshbach, Villars and other authors, it is concluded that the even part of an operator (but not the operator of non-local theory) is the observable.

In the case of a free particle, the even part of the position operator coincides with the one in the non-local theory (in this case, the Newton—Wigner position operator). The momentum operator does not include an odd part.

One has another situation in the case of a particle in a constant homogeneous magnetic field. It is shown that in this case the even parts of the operators which describe the rotational motion generate the deformed Heisenberg—Weyl algebra. Moreover, they do not commute with the relevant “mean position” operator for the motion along the field.

The question about the matrix-valued Wigner function formalism for scalar charged particles in the Feshbach and Villars’ approach is considered in *THIRD CHAPTER* “Representation of matrix-valued symbols”. The general results are the writing down of the relevant Weyl transformation and the fact that under some conditions matrix-valued Moyal bracket does not coincide with the matrix-valued Poisson bracket in the classical limit.

In *FOURTH CHAPTER* “Phase space representation for charge-invariant observables in the case of a free particle”, consideration is restricted to observables that have matrix-valued Weyl symbols proportional to the identity matrix. These observables can be written in the following form:

$$A_\alpha{}^\beta(p,q) = A(p,q)\delta_\alpha{}^\beta .$$

In a fact, such observables are combinations of the position and momentum. It is shown that the even and odd parts of such operators are uniquely related to each other.

This relation is a consequence of the fact that the expressions for even and odd parts of the position operator are known and have fixed values.

It is possible to introduce the standard (not matrix-valued) Wigner function for charge-invariant observables in such a way that their means are defined by the standard rule:

$$\overline{A} = \int_{-\infty}^{+\infty} A(p,q)W(p,q)dpdq .$$

This object includes four components: two even ones

$$W_{[\pm]}(p,q)=\frac{1}{(2\pi\hbar)^d}\int_{-\infty}^{+\infty}\varepsilon\left(p+\frac{P}{2},p-\frac{P}{2}\right)\psi_{\pm}^{*}\left(p+\frac{P}{2}\right)\psi^{\pm}\left(p-\frac{P}{2}\right)\exp\left(-\frac{i}{\hbar}Pq\right)dP\,,$$

and two odd ones

$$W_{\{\pm\}}(p,q)=\frac{1}{(2\pi\hbar)^d}\int_{-\infty}^{+\infty}\chi\left(p+\frac{P}{2},p-\frac{P}{2}\right)\psi_{\pm}^{*}\left(p+\frac{P}{2}\right)\psi^{\mp}\left(p-\frac{P}{2}\right)\exp\left(-\frac{i}{\hbar}Pq\right)dP\,,$$

where $\varepsilon(p_1,p_2)$, $\chi(p_1,p_2)$ are the functions expressed via energy spectrum. They play a crucial role in the consideration presented, and are called as $\varepsilon$-and $\chi$-factors.

The physical meaning can be ascribed only to the even components. The odd components differ from zero only for hypothetical states which are superpositions of a particle and anti-particle. The existence of such states is prohibited by the charge superselection rule. However, the odd part plays a crucial role in processes with unstable vacuum, and thus is considered as well in the thesis.

The definition of the even part of the Wigner function differs from the one in the non-local (or, non-relativistic) theory. $\varepsilon$-factor describes the influence of the non-trivial vacuum structure on the form of Wigner function.

However, it turned out that the evolution equation for the even part of the Wigner function coincides with the one in the non-local theory and can be written as follows:

$$\partial_t W_{[\pm]}(p,q,t)=\{E(p),W_{[\pm]}(p,q,t)\}_M\,,$$

where $E(p)=\sqrt{m^2c^4+c^2p^2}$ .

So-called “Anti-Moyal” bracket (the symbol analogue of the anticommutator) appears in the equation for the odd part of the Wigner function:

$$\partial_t W_{\{\pm\}}(p,q,t)=\lfloor E(p),W_{\{\pm\}}(p,q,t)\rfloor_M\,.$$

In this case, the difference between the standard and non-local theories is connected with the fact that functions, that can present real physical states, belong to different classes. It is particularly visualized for the criteria of pure state. For the even part of the Wigner function the relevant condition can be written as follows:

$$\frac{\partial^2}{\partial p_1\partial p_2}\ln\int_{-\infty}^{+\infty}W_{[\pm]}\left(\frac{1}{2}(p_1+p_2),q\right)\exp\left(\frac{i}{\hbar}(p_1-p_2)q\right)dq=-\frac{c^4p_1p_2}{E(p_1)E(p_2)(E(p_1)+E(p_2))^2}\,.$$

Unlike the non-local (non-relativistic) theory, the right-hand side of the expression is not zero.

In particularly, it results in impossibility of existence of state given by the Gauss distribution for the both position and momentum. The contours of the Wigner function for the state that is the Gauss distribution near the zero point in the momentum space are shown in Fig. 1. The characteristic size of this package is about 8 times less than the Compton wavelength. The very strong localization leads to appearance of specific “vacuum fluctuations”. These are the reason that the second moment (dispersion) has some peculiarities in the standard theory. For example, the dispersion of the state shown in Fig. 1 is negative. However, from the other hand the main part of the Wigner function is within the characteristic length.

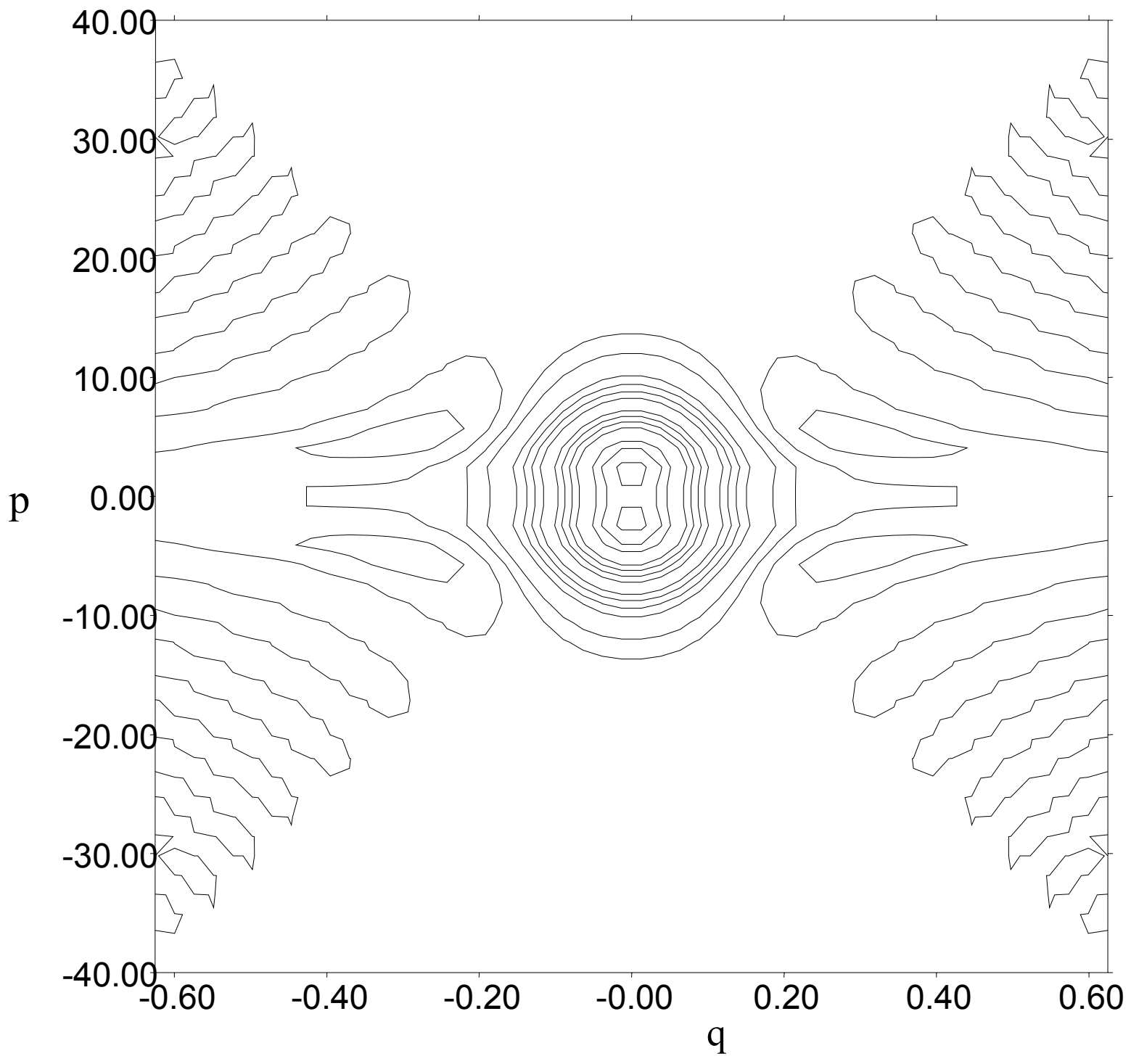

Fig.1. Contours of the Wigner functions for the "Gauss" state of a free particle. The localization parameter is $\lambda = 8$. Position is given in $\lambda_c = \hbar / mc$ units, momentum is given in $mc$ units.

The same consideration but for a more complicated case is given in *FIFTH CHAPTER* "Phase space representation for charge-invariant observables in the case of a particle in a constant homogeneous magnetic field".

Here, quantum dynamics differs from classical one more essentially. The reason is the fact that the classical Hamilton function does not coincide with the symbol that plays role of the Hamiltonian in the consideration presented. The relativistic square root is determined by means of the star-product. This leads to the fact that localization parameter $\lambda$ (ratio of the Compton wavelength to the characteristic length) serves as another relativistic parameter along with velocity. It is worth noting that this feature is common for both the standard and non-local theories due to the equivalence of the evolution equations. This is similar to the case of a free particle considered in the previous Chapter.

In this Chapter the question about the physical meaning of the $\varepsilon$-factor is considered. It is an additional multiplier for the interference terms between eigenstates of the Hamiltonian. Therefore, the non-trivial vacuum structure leads to effective increase of coherence. Assume that a scalar chargeparticle is considered as an open system that along with its environment is in such the entangled state that each eigenstate of the Hamiltonian is related to a macroscopically distinct state of the environment. Then the reconstruction of the Wigner function (by means, for example, quantum tomography method) gives more information about the interference terms. It is worth noting that the coherence does not increase here because it does not influence any of quantitative characteristics of the entanglement. The vacuum structure plays the role of so-called "quantum lens": in such kind of measurements, the coherence seems to increase for the observer (he gets more evident information about the interference terms) but in fact it is not true.

*SIXTH CHAPTER* "Coherent states of a relativistic particle" is devoted to consideration of such coherent states which, first of all, do not use the zero plane concept, and, second, satisfy the following additional conditions:

1. The means of *standard* position and momentum are connected with the parameter $\alpha$ in a usual way:

$$\alpha = \frac{1}{\sqrt{2}}\left(\frac{\overline{q}}{\sigma} + i\frac{\sigma}{\hbar}\overline{p}\right)$$

2. These states are expanded in series by eigenstates of the Hamiltonian which have the same sign of charge.

At first sight, these conditions are incompatible. Indeed, to satisfy the first, one should define the coherent states as, for example, eigenstates of the standard annihilation operator. However, it has the non-trivial charge structure, and the second condition is not satisfied in this case. On the other hand, one can define coherent states using the annihilation operator of non-local theory. Then, another situation takes place: the second condition is satisfied, but the first one is not.

It is easy to find a way from this situation. To satisfy the both conditions one should define the coherent states as eigenstates of *the even part of the standard annihilation operator*.

One can emphasize two groups of the peculiarities related to this approach. First, it is deviation of the wave packages trajectories from classical ones following from the fact that the Hamiltonian is not quadratic. Second, these are "vacuum fluctuations" following from the non-trivial charge structure of the position operator.

In the case of a free particle the first group of effects results in the effective increase of mass for a strongly localized particle even for small (non-relativistic) velocities. This is a consequence of the fact that when the position dispersion is small, the momentum dispersion becomes large, and relativistic values of the momentum contribute to the general picture of motion. This effect is similar to the effective increase of mass (inertia) at large velocities in special relativity. However, in the quantum case, the localization parameter is also relativistic.

For the rotational motion of a particle in a constant homogeneous magnetic field this peculiarity leads to a low frequency modulation of the orbit radius. It is worth noting, that such effects are absent in approaches using the zero plane concept. Therefore, if such peculiarities were observed at least in the synchrotron radiation from astrophysical objects alike neutron stars, this would indirectly indicate the reference frame where reduction of a quantum state takes place.

The second group of the effects (related to the vacuum structure) almost does not manifest itself for the first moments of the position and momentum. One can see some peculiarities for relativistic rotator (here we deal with non-linear coherent states) in the case of a strong localization and small orbit radius only.

However, this manifests itself very well for the second moments (dispersions). In the case of a particle with very strong localization, "vacuum fluctuations" are concentrated near the zero value of the momentum. Thus, when its values are large enough, a particle "senses" this fluctuation weakly. As that, the behavior of the position dispersion is similar to that in the non-local theory.

The situation is opposite in the case of relativistic rotator. The state with zero orbit radius is the ground state for this system. It does not have any features of the vacuum structure. For sufficiently large radius, superposition of the eigenstates with the interference terms appears. The additional multipliers ($\varepsilon$-factor) lead to appearance of "vacuum fluctuations".

Non-commutability of the even parts of the annihilation operators for the rotational and transnational motions makes it impossible to construct such states for the both degrees of freedom. One can introduce coherent states that describe one of the two degrees assuming only a finite localization along the other one. This is done in this Chapter.

The influence of relativistic effects (namely, non-locality of the Hamiltonian in the Feshbach—Villars or Foldy—Wouthuysen representations) on energy characteristic of entangled coherent states is considered in the *SEVENTH CHAPTER* "Entangled coherent states for free relativistic particles".

Symmetry (anti-symmetry) of the wavefunction results in appearance of a specific correlation term in the mean energy. In the case of scalar charged particles it does not lead to essential peculiarities. This is explained by the fact that two such particles can be in the same state simultaneity.

However for fermions, the state when two particles are in the same point is peculiar. The reason is the Pauli exclusion principle. When one brings together the wave packages, one of the particles transits to the first excited displacement number state. The energy of this system is larger than the energy of two separated particles. Thus, generally speaking, fermions save energy keeping as far as possible from each other (the phenomena of Fermi repulsion).

This feature somewhat changes due to non-locality of the relativistic Hamiltonian in the Foldy—Wouthuysen representation. Namely, the energy difference of two packages localized over the Compton wavelength when they are centered at the same point and when they are well separated, becomes smaller. This means that such a state is more energetically favored, and thus has more chances to be realized as a result of a physical process.

The general results of the thesis and recommendations for their applications are given in *CONCLUSIONS*.

# CONCLUSIONS

The thesis presents theoretical consideration of the scientific problem, which appears as localization of a scalar charged particle in the phase space representation. It is considered how the quantum relativistic effects can be described in the phase space representation. Namely, a particular attention is paid to the vacuum structure and effective non-locality of the Hamiltonian. Some new physical effects are predicted theoretically, and a possibility of their applications in fundamental and applied studies is discussed.

In the case of a free particle the even part of the position operator coincides with the Newton—Wigner position operator. Along with the standard momentum operator (in this case it does not contain the even part), they are the standard canonical pair. The situation is quite different when a constant magnetic field is present in the system. The even parts of the position and momentum operators (creation and annihilation) which describe the rotational motion do not coincide with the relevant operators of the non-local theory. Moreover, they present a deformation of the last. Therefore, the "mean positions" operators generate the deformed Heisenberg—Weyl algebra with the correspondent commutation relations. Furthermore, the relevant operator for the translational motion is shown to not commutate with them. This means, that a particle can not be found in a state which is exactly described by the both: translational and rotational positions.

The consideration of the WWM formalism is restricted to observables which are arbitrary combinations of the standard position and momentum and do not depend on the charge variable (charge-invariant observables). In other words, one can say that the matrix-valued Weyl symbols of such observables are proportional to the identity matrix. The even and odd parts of such observables are shown to be uniquely related to each other. This relation has an evident physical meaning. For instance, if one takes the scalar potential of an external electric field as an example of such operator, this relation describes the quantitative connection between the motion of the particle in the field and the effects related to polarization of the vacuum.

It is possible to introduce the standard (not matrix-valued) Wigner function for charge-invariant observables. This object includes four components: two even ones and two odd ones. The odd part differs from zero for hypothetical states that are superposition of a particle and antiparticle. The general physical meaning can be ascribed to the even components only. The evolution equations for them are the same as those in the non-local theory.

There is a difference in the definition of the even components compared to the non-relativistic case and non-local theory. This is related to the existence of the specific function of two variables, $\varepsilon$-factor. Its physical meaning is the additional multiplier for interference terms between eigenstates of the Hamiltonian. It is very important, that the $\varepsilon$-factor exceeds unity. This means an effective increase in the coherence in such systems (or, to be more precise, in such kinds of measurements).

If the particle and an environment are in such an entangled state that every eigenstate of the Hamiltonian is related to certain macroscopically distinct state of the environment, information about the interference terms (relative phase) is lost due to the decoherence processes. Hence, if one reconstructs the Wigner function in an experiment (for example, using the quantum tomography method), their influence will be quite invisible. The effective increase in coherence plays the role of a quantum lens in this case: though these terms do not increase objectively, they became more visible for an observer as a result of measurement. This effect can be applied to study decoherence, when one needs to reconstruct very small interference terms.

Such effects can also be observed in other systems with band structure of the energy spectrum, whose study has nowadays become a real task for the modern experimental technics. Hence, it is possible to apply these for the decoherence suppression in the quantum computers. The formalism presented can be used for developing relativistic quantum tomography methods and tomography of the conduction electrons in solids.

The relativistic coherent states that take into account non-trivial charge structure of the position and momentum and satisfy the charge superselection rule are constructed in the thesis. To satisfy both conditions one can define the coherent states as eigenstates of the even part of the annihilation operator.

A low frequency modulation of the orbit radius for a particle in a constant homogeneous magnetic field is found in this approach. It is worth noting that this effect is absent in the approach based on the concept of the zero plane. Hence, it can be quite a good test for determination of the reference frame where the quantum state reduction takes place (indeed, under the condition that this can be observed, for example, in the synchrotron radiation from astrophysical objects similar to neutron stars).

The influence of the entanglement on the energy means has been considered in the thesis as well. These include the correlation energy as a specific term in addition to the energy of separate particles. Scalar charged particles obey the Bose—Einstein statistic, thus the correlation energy does not play a crucial role here. However, the presented formalism is well adopted for the case of one-dimensional Dirak particles. The Fermi repulsion of two such particles results in an excitation one of them the first displacement number state when both particles are localized at the same point of the phase space. At that, the mean energy of this state is lager than the energy of two spatially separated particles. Therefore, energywise this state is not favorable, and a large interparticle separation has a higher probability. Non-locality of the Hamiltonian in the Foldy—Wouthuysen representation brings same changes here. For an entangled coherent state localized over the Compton wavelength, this energy decreases, and this state becomes of higher probability.

## LIST OF PUBLICATIONS

## ABSTRACT

Semenov A. A. Phase space localization of a scalar charged particle. – Manuscript.

Thesis for the degree of Doctor of Philosophy (Candidate of Physics and Mathematics) by speciality 01.04.02 – Theoretical Physics. – Physics Department, Kiev National Taras Shevchenko University, Kiev, 2002.

The thesis is devoted to the phase space representation of relativistic quantum mechanics. For a class of observables with matrix-valued Weyl symbols proportional to the identity matrix, the Weyl—Wigner—Moyal formalism is proposed. The evolution equations are found to coincide with their counterparts in relativistic quantum mechanics with non-local Hamiltonian. The difference between the theories is connected with peculiarities of the constraints on the initial conditions. Effective increase in coherence between eigenstates of the Hamiltonian is found.

Relativistic coherent states that take into account a non-trivial charge structure of the position and momentum operators and satisfy the charge superselection rule are considered. On this basis, the entangled coherent states are developed.

# АНОТАЦІЯ

Семенов А.О. Локалізація скалярної зарядженої частинки у фазовому просторі. – Рукопис.

Дисертація на здобуття наукового ступеня кандидата фізико-математичних наук за спеціальністю 01.04.02 – теоретична фізика. – Київський національний університет ім. Тараса Шевченка, фізичний факультет, Київ, 2002.

В дисертаційній роботі вивчається представлення релятивістської квантової механіки на фазовому просторі. Для класу спостережуваних, матричнозначні символи Вейля яких пропорційні одиничній матриці, запропонований формалізм Вейля – Вігнера – Мояла. Еволюційні рівняння співпадають із своїми аналогами в нелокальній теорії, а відмінності виявляються в особливостях обмежень для можливого класу початкових умов. Знайдене ефективне подавлення декогерентності між власними станами гамільтоніана.

Розглянуто релятивістські когерентні стани, що одночасно враховують нетривіальну зарядову структуру операторів координат та імпульсів, і задовольняють правилу супервідбору. На цій основі представлений розгляд переплутаних когерентних станів релятивістської частинки.

# АННОТАЦИЯ

Семенов А.А. Локализация скалярной заряженной частицы на фазовом пространстве.-Рукопись.

Диссертация на соискание ученой степени кандидата физико-математических наук по специальности 01.04.02 – теоретическая физика. – Киевский национальный университет им. Тараса Шевченко, физический факультет, Киев, 2002.

В диссертационной работе изучается представление релятивистской квантовой механики на фазовом пространстве. Для класса наблюдаемых, чьи матричнозначные символы Вейля пропорциональны единичной матрице, предложен формализм Вейля – Вигнера – Мояла. Эволюционные уравнения совпадают со своими аналогами в нелокальной теории, а отличия проявляются в особенностях ограничений на возможный класс начальных условий. Обнаружено эффективное подавление декогерентности между собственными состояниями гамильтониана.

Рассмотрены релятивистские когерентные состояния, которые одновременно учитывают нетривиальную зарядовую структуру операторов координат и импульсов, и удовлетворяют правилу суперотбора. На этой основе представлено рассмотрение перепутанных когерентных состояний релятивистской частицы.